\documentclass[prd,twocolumn,preprintnumbers,amsmath,amssymb,nofootinbib]{revtex4-2}

\usepackage{array}
\usepackage{multirow}
\usepackage[T1]{fontenc}
\usepackage[utf8]{inputenc}
\usepackage{microtype}
\usepackage{graphicx}
\usepackage{booktabs}
\usepackage{hyperref}
\usepackage{siunitx}
\usepackage{rotating}
\usepackage{comment}
\usepackage{tabularx}
\usepackage{array}
\newcolumntype{Y}{>{\raggedright\arraybackslash}X}

\usepackage{xspace}

\hypersetup{colorlinks=true,linkcolor=blue,citecolor=blue,urlcolor=blue}

\newcolumntype{Y}{>{\raggedright\arraybackslash}X}
\newcommand{\gbm}{Fermi/GBM\xspace}
\newcommand{\swift}{\textit{Swift}\xspace}
\newcommand{\lat}{Fermi/LAT\xspace}
\newcommand{\konus}{Konus-Wind\xspace}

\newcommand{\ipn}{IPN\xspace}

\begin{document}

\title{Backwards Gamma-Ray Bursts: Searching for Exploding Primordial Black Holes in Short-Duration GRB Catalogs}

\author{Stefano Profumo}
\affiliation{Santa Cruz Institute for Particle Physics (SCIPP), Santa Cruz, CA 95064, USA}
\affiliation{University of California, Santa Cruz, CA 95064, USA}
\author{Kally Wen}
\affiliation{Lynbrook High School, San Jose, CA 95129, USA}

\date{\today}

\begin{abstract}
\noindent We present a systematic search for signatures of terminal black-hole evaporation in short gamma-ray burst (sGRB) catalogs. An exploding primordial black hole (PBH) undergoing final-stage Hawking radiation is predicted to produce a distinctive ``backwards burst''---a very short, spectrally hard transient with monotonically increasing flux and little or no longer-wavelength afterglow. We develop a forward-modeling framework that directly compares theoretical PBH evaporation light curves, computed with full Standard Model particle content and detector response folding, against empirical GRB pulse templates. Analyzing 39 well-characterized {\it Swift} sGRBs with non-detected or extremely faint afterglows, we find that all events exhibit fast-rise, slow-decay temporal profiles inconsistent with the PBH prediction. Model comparison via Akaike and Bayesian information criteria decisively favors conventional FRED or ERCA fits over the PBH template for every burst. No candidates for terminal PBH evaporation are identified. The null result yields an upper bound on the local PBH explosion rate density $R_{\mathrm{PBH}} \lesssim 10^5~\mathrm{pc}^{-3}~\mathrm{yr}^{-1}$, comparable to constraints from dedicated TeV $\gamma$-ray searches. Our methodology establishes a robust template-matching approach that can be scaled to larger multi-instrument catalogs, providing a foundation for future searches targeting this unique signature of quantum gravity and early-Universe physics.

\end{abstract}

\maketitle

\section{Introduction}
Black-hole evaporation via quantum particle creation predicts that sufficiently low-mass black holes radiate thermally and ultimately ``explode'' in a runaway final phase~\cite{Hawking1974,Hawking1975}. 
In the standard semiclassical treatment, a non-rotating, uncharged black hole of mass $M$ emits a nearly blackbody spectrum of all kinematically accessible particle species at the Hawking temperature $T_{\rm H}\propto M^{-1}$, losing mass at a rate $\dot M\propto -M^{-2}$ and evaporating on a timescale $t_{\rm evap}\propto M^{3}$. 
Black holes with initial masses below a critical value $M_\star\sim (5$--$8)\times 10^{14}\ {\rm g}$ therefore have lifetimes shorter than the age of the Universe and would be completing their evaporation today. 
If such light black holes formed in the early Universe---\emph{primordial} black holes (PBHs)---their terminal evaporation would produce very-short, very-hard bursts of high-energy radiation in the local cosmos.

PBHs are a well-motivated consequence of large-amplitude primordial perturbations and other non-standard early-Universe phenomena, and they remain an actively explored dark-matter candidate across a broad mass range~\cite{Carr:2016drx,Carr:2020gox,Villanueva-Domingo:2021spv}.  
For masses near $M_\star$, Hawking evaporation provides both a powerful constraint and a unique discovery channel: the cumulative emission of a cosmological PBH population contributes to the diffuse $\gamma$-ray and cosmic-ray backgrounds, while the final explosive phase of nearby objects would manifest as individual transients. 
Diffuse $\gamma$-ray measurements with \textit{CGRO}/EGRET and \textit{Fermi}-LAT, as well as charged-particle measurements, already place stringent bounds on the cosmological abundance of light PBHs~\cite{Carr:2016drx, Korwar:2023kpy, Korwar:2024ofe}. 
Complementary searches for short bursts in TeV $\gamma$ rays with Milagro, HAWC, and H.E.S.S.\ constrain the local rate density of PBH explosions, but still allow rare events within the Solar neighborhood~\cite{Abdo:2015osa,Albert:2019lkt,HESS:2021pbh}. 
A dedicated \textit{Fermi}-LAT analysis of steady emission from nearby PBHs likewise finds no evidence but leaves open a wide range of possibilities for the local PBH distribution~\cite{FermiLAT:2018pbh}. 
From a particle-physics perspective, the detection of even a single evaporating PBH would be transformative: the detailed spectrum and time evolution of Hawking radiation can, in principle, probe the full particle content of nature---including dark-sector degrees of freedom that couple only gravitationally~\cite{Baker:2021sno,Arbey:2019cmf, Boluna:2023jlo, Federico:2024fyt, Ewasiuk:2024ctc}.

The expected transient associated with the terminal evaporation of a PBH has been discussed since the early days of GRB astronomy~\cite{MacGibbonCarr1991,Belyanin:1996,Belyanin:1998}. 
As the black hole mass decreases, both the luminosity and the characteristic photon energy increase rapidly, so that in the rest frame of the explosion the light curve is highly asymmetric: the emission is faint and relatively soft at early times, and culminates in a brief, extremely hard spike just prior to complete evaporation. 
When expressed as a function of ``remaining time to evaporation,'' the profile is therefore strongly rising, with a characteristic spectral hardening toward the end. 
This behavior is the opposite of the canonical hard-to-soft evolution seen in many GRB pulses, motivating the colloquial notion of ``backwards'' GRBs as a signature of PBH explosions.
Early studies using BATSE data searched for very short GRBs (VSGRBs) with durations $T_{90}\lesssim 100~{\rm ms}$ and unusually hard spectra, and argued that such events might be consistent with PBH evaporation~\cite{Cline:1995,Cline:1996,Cline:2007pbh}. 
Subsequent work, however, has shown that the VSGRB population can largely be accommodated within the broader short-GRB phenomenology or attributed to other compact-object or magnetar-related progenitors, and robust PBH candidates have not emerged from these searches~\cite{Carr:2016drx}.

Gamma-ray bursts themselves exhibit a well-known bimodality in duration, with ``short'' GRBs typically defined by $T_{90}\lesssim 2$~s and harder prompt spectra than long bursts~\cite{Kouveliotou1993}. 
Multiwavelength follow-up and host-galaxy studies now firmly link the majority of short GRBs to compact-object mergers, most commonly binary neutron stars and neutron star--black hole systems~\cite{Nakar2007,Berger2014}. 
In the standard picture, the prompt high-energy emission is followed by a long-lived afterglow produced as a relativistic outflow decelerates in the circumburst medium, giving rise to broadband synchrotron radiation from X-ray to radio frequencies~\cite{SariPiranNarayan1998}. 
A sizable fraction of short GRBs exhibit X-ray and optical afterglows, and for events with detailed follow-up one often finds extended emission, plateaus, or flares that further support the external-shock and central-engine interpretations~\cite{Dichiara2020}. 
From the perspective of PBH explosions, this is a crucial qualitative difference: a terminal evaporation event is not expected to drive a long-lived external shock, and may therefore lack a detectable afterglow altogether, apart from any emission powered directly by the Hawking flux in the immediate vicinity of the black hole.

The growing sample of short GRBs with deep, systematic follow-up has revealed a minority sub-population with either non-detections or extremely faint limits on afterglow emission. 
In particular, Dichiara et al.~\cite{Dichiara2020} compiled a carefully vetted catalog of \swift\ short bursts within $\sim 200$~Mpc and categorized them according to the quality and outcome of the afterglow searches. 
This population, which spans a range of prompt durations and spectral properties, naturally invites consideration of more exotic progenitors, such as off-axis or choked jets, magnetar giant flares, or PBH explosions. 
In this context, the PBH ``backwards burst'' hypothesis suggests a concrete operational discriminant: candidate events should be very short and spectrally hard, \emph{and} should lack an X-ray afterglow despite rapid, sensitive follow-up.

In this work we revisit the PBH--GRB connection by constructing a forward-modeling framework that confronts realistic PBH evaporation light curves with the observed population of short GRBs without detected afterglows. 
Using state-of-the-art calculations of Hawking emission, including grey-body factors and secondary particle production, we generate time- and energy-resolved photon spectra for PBHs nearing their terminal evaporation, and propagate these through the \swift\ detector response to obtain instrument-specific count light curves. 
We then perform a systematic comparison between this PBH ``backwards burst'' template and standard phenomenological models of GRB temporal profiles, using the Dichiara et al.\ sample of short \swift\ bursts with non-detected afterglows as our data set. 
Our goal is twofold: (i) to assess whether any events in this well-characterized sample are consistent with the expectations for terminal PBH evaporation, and (ii) to establish a methodology that can be scaled up to larger GRB catalogs and other instruments, thereby sharpening future searches for exploding black holes in high-energy transient data. As a by-product, we will also (iii) derive a limit on the local PBH explosion rate density.

\section{Short GRBs and “backwards” GRBs}
\label{sec:shortgrbs}

Gamma-ray bursts (GRBs) exhibit a well-established bimodality in duration and spectral hardness~\cite{Kouveliotou1993}. 
The short-duration population ($T_{90}\!\lesssim\!2$ s) typically shows harder prompt spectra and negligible spectral lags relative to the long-duration class, which is associated with the core collapse of massive stars. 
Since the advent of multiwavelength follow-up, the progenitors of most short GRBs (sGRBs) have been securely linked to compact-object mergers---binary neutron star or neutron star--black hole systems---whose relativistic jets produce brief, intense $\gamma$-ray flashes followed by broadband synchrotron afterglows as the outflow decelerates in the ambient medium~\cite{Nakar2007,Berger2014}. 
Extended emission, plateaus, and X-ray flares seen in a subset of short bursts provide further evidence for central-engine activity and external-shock afterglows~\cite{Kaneko2015,Dichiara2020}.

A terminal primordial black-hole (PBH) evaporation event, by contrast, is fundamentally different in both physical origin and temporal morphology.  
In the semiclassical picture of Hawking radiation, the evaporation rate accelerates as the black-hole mass decreases, yielding an effective luminosity $L\!\propto\!M^{-2}$ and temperature $T_{\rm H}\!\propto\!M^{-1}$.  
The instantaneous photon flux therefore scales approximately as $\dot N_{\gamma}\!\propto\!t_{\mathrm{rem}}^{-p}$, with $t_{\mathrm{rem}}$ the remaining time to evaporation and $p\simeq0.3$–0.4 depending only weakly on the particle content and energy band considered~\cite{MacGibbonCarr1991,Arbey:2019cmf}.  
Because these scalings follow directly from the universal mass–temperature relation, the {\it shape} of the final light curve is effectively universal for all nonrotating PBHs, modulo redshift and instrumental response.  
This universality underlies the concept of a “standard” PBH explosion template: the emission steeply rises and hardens with decreasing $t_{\mathrm{rem}}$, peaking in the final $\sim10^{-2}$–$10^{-1}$ s, and terminating abruptly once the black hole reaches the Planck regime.  
Expressed in the usual time coordinate $t$, the light curve appears inverted relative to canonical GRB pulses—rising slowly and decaying rapidly when viewed forward in time—hence the designation of “backwards bursts.”

The predicted spectral evolution also provides a key discriminant.  
Whereas most astrophysical GRB pulses exhibit {\it hard-to-soft} evolution, with the $\nu F_\nu$ peak energy $E_{\mathrm{p}}$ decreasing through the pulse, the PBH evaporation spectrum evolves in the opposite sense: $E_{\mathrm{p}}$ and the overall hardness ratio both {\it increase} monotonically as the remaining mass decreases~\cite{Belyanin:1996,Belyanin:1998,Cline:1995,Cline:1996,Cline:2007pbh}.  
Consequently, the joint temporal–spectral behavior of a PBH burst is expected to be distinctive: an ultra-short, hard transient whose photon index softens only at the very end, if at all, and whose cumulative fluence is heavily dominated by the final milliseconds.

Observationally, the compact-binary-merger scenario and the PBH-evaporation hypothesis occupy complementary corners of parameter space.  
Typical sGRBs last $\sim0.1$–$2$ s, with measurable afterglows and often host-galaxy associations; PBH explosions would appear as isolated, sub-second flashes—durations $T_{90}\!\sim\!10$–$200$ ms—without an accompanying supernova, host, or afterglow.  
In practice, this motivates a phenomenological search criterion: identify events with (i) very short and hard prompt emission, (ii) no X-ray or optical afterglow despite rapid, deep follow-up, and (iii) no evidence of an extended emission phase or magnetar-like tail.  
Such bursts, if genuine and non-instrumental, would constitute plausible PBH-candidate “backwards” GRBs.

Finally, the universality of the PBH light curve allows for robust forward-modeling and statistical comparison across instruments.  
Once folded through a detector’s energy response, the normalized PBH count profile depends primarily on two free parameters—the total fluence (or distance) and the absolute onset time—making it an ideal template for matched-filter searches in archival GRB data.  
By contrast, empirical models such as the fast-rise–exponential-decay (FRED) or exponential-rise–constant-afterglow (ERCA) forms typically require multiple shape parameters ($\tau_1$, $\tau_2$, $t_s$, etc.).  
As we show below, these differences enable quantitative tests of the PBH hypothesis in well-characterized sGRB samples.

\section{Data Selection and Catalogs}
\label{sec:data}

A meaningful search for PBH evaporation signatures in the short–duration gamma-ray burst population requires a well-defined, homogeneous dataset with reliable prompt-emission characterization and uniformly constrained afterglow follow-up.  
Our goal is to isolate events that could, in principle, be consistent with the ``backwards-burst'' light curve expected from a terminal PBH evaporation—namely, ultra-short, spectrally hard transients with no persistent emission at longer wavelengths.

\subsection{Baseline selection: short GRBs without detected afterglows}

We begin from the curated sample assembled by Dichiara et al.~\cite{Dichiara2020}, who analyzed all \swift\ short GRBs ($T_{90}\!\lesssim\!2$ s) detected between 2005 and 2019 and systematically classified them according to the outcome of their multiwavelength afterglow searches.  
This catalog provides uniform burst parameters, localization uncertainties, and upper limits on X-ray, optical, and radio afterglows obtained with \swift/XRT, UVOT, and ground-based telescopes.  
From this parent sample we select the subset of events with \textit{no detected afterglow} or with follow-up upper limits stringent enough to rule out a typical external-shock afterglow at distances $\lesssim200$ Mpc.  
This ensures that the absence of afterglow emission reflects physical faintness rather than incomplete follow-up.

\subsection{Cross-matching and multi-instrument validation}

For each candidate we cross-match with other major GRB instruments to ensure prompt detection and to extract additional temporal and spectral information.

\paragraph*{{\it Fermi}/GBM and LAT.}
The \gbm\ catalog~\cite{Meegan2009} provides wide-field, nearly continuous coverage of the unocculted sky in the 8 keV–40 MeV range.  
For each \swift\ burst we check for onboard GBM triggers and ground-search detections in the time window $\pm30$ s around the \swift\ trigger.  
When available, we use the GBM time-tagged event (TTE) data to confirm the light-curve morphology and extract the prompt spectral parameters (peak energy $E_{\mathrm{p}}$, photon index, fluence).  
We also check for high-energy counterparts in the \lat\ catalog~\cite{Atwood2009}, although none of the events considered here shows significant $>100$ MeV emission.

\paragraph*{Konus-Wind.}
The \konus\ experiment~\cite{Aptekar1995} has operated continuously since 1994 and provides independent coverage of bright GRBs in the 20 keV–15 MeV band from an interplanetary vantage point.  
Its light curves and spectral fits are invaluable for confirming short, intense bursts and for ruling out solar or magnetospheric particle events.  
Where possible, we cross-reference each candidate against Konus trigger lists and published analyses.

\paragraph*{INTEGRAL.}
INTEGRAL's instruments (SPI-ACS and IBIS) contribute complementary high-time-resolution coverage of bright short bursts \cite{Winkler2003}.  
Although not spectrally resolved, SPI-ACS detections provide robust confirmation of astrophysical transients and enable temporal cross-checks across instruments.

\paragraph*{Interplanetary Network (IPN).}
The Interplanetary Network (\ipn)~\cite{Hurley2013,Palshin2013,HEASARCIPN} combines detectors aboard multiple spacecraft distributed throughout the Solar System, including \textit{Wind}/Konus, \textit{Mars Odyssey}/HEND, and \textit{INTEGRAL}/SPI-ACS.  
By triangulating burst arrival-time differences, the \ipn\ provides annular localization constraints with typical widths of tens of arcminutes.  
An \ipn\ detection thus serves as both confirmation of astrophysical origin and a means of refining sky positions for host-galaxy searches. Critically for the present work, IPN offers the opportunity to constrain the proximity of the event to within a distance compatible with the detectability of a PBH evaporation event \cite{Boluna:2023jlo, Federico:2024fyt}.

\subsection{Event verification and contamination control}

Each candidate event is screened for possible contamination by non-GRB transients, such as magnetar giant flares, solar-particle events, or instrumental artifacts.  
We adopt the event-type classifications provided in the literature (e.g., Dichiara’s Groups a–d) and flag potential contaminants, but retain them in our tables for completeness.  
Cross-instrument consistency, localization agreement, and the presence of a characteristic GRB-like temporal profile are required for inclusion in the final working list.

\subsection{Restricted sample selection}

Although much larger GRB catalogs exist (including \gbm\ and \ipn\ samples numbering thousands of events), several considerations motivate focusing on the \swift\ short-burst subset with deep afterglow constraints:

\begin{itemize}
    \item \textbf{Uniform follow-up:} \swift's rapid autonomous slewing enables near-real-time XRT and UVOT observations, ensuring consistent upper limits or detections across the sample—essential when the absence of an afterglow is the discriminant.
    \item \textbf{Precise localizations:} \swift/BAT positions are accurate to a few arcminutes, permitting host-galaxy association or exclusion; this is not possible for GBM-only events with multi-degree error regions.
    \item \textbf{Homogeneous temporal resolution:} BAT light curves (15–350 keV) provide sub-millisecond timing, sufficient to resolve the rapid evolution expected from a PBH explosion.
    \item \textbf{Cross-instrument validation:} Many \swift\ bursts are simultaneously detected by GBM or Konus-Wind, allowing for spectral cross-checks and rejection of instrumental artifacts.
\end{itemize}

These criteria balance sensitivity, completeness, and data quality.  
Restricting to well-characterized short bursts minimizes false positives due to incomplete follow-up, while retaining enough events to enable statistical statements.  
The resulting working sample, summarized in Table~\ref{tab:list}, serves as the foundation for our temporal-profile analysis in the following sections.

\begin{sidewaystable*}
\centering
\scriptsize
\setlength{\tabcolsep}{4pt}
\renewcommand{\arraystretch}{1.15}
\caption{Short GRBs with undetected afterglows (following Ref.~\cite{Dichiara2020}) and ancillary flags used in this study.
``Grp'' and ``Reason'' reproduce the grouping and rationale in Ref.~\cite{Dichiara2020}. 
``{\it Fermi}/GBM LC'' indicates availability of a \gbm\ prompt light curve (triggered or sub-threshold). 
``Prompt spectrum'' notes whether a published time-averaged prompt spectral fit is available (from \swift/BAT or \gbm). 
``IPN'' flags membership in the Interplanetary Network.}
\label{tab:list}
\begin{tabular}{@{}l c l l l c@{}}
\toprule
GRB & Grp & Reason & {\it Fermi}/GBM LC & Prompt spectrum & IPN \\
\midrule
GRB 080121B & a & likely missed due to very short/low-fluence prompt emission &
No – pre-\textit{Fermi} &
Likely via \swift/BAT refined analysis (GCN) &
No evidence \\
GRB 090815C & a & likely missed due to very short/low-fluence prompt emission &
Yes – triggered (GBM) &
Yes ({\it Fermi}/GBM catalog \& GCN) &
No evidence \\
GRB 091117 & a & likely missed due to very short/low-fluence prompt emission &
No record in GBM catalogs/GCNs &
Likely via \swift/BAT refined analysis (GCN) &
No evidence \\
GRB 100216A & a & likely missed due to very short/low-fluence prompt emission &
Yes – subthreshold/ground (GBM) &
Possible (GBM targeted/ground search; check CTTE products) &
No evidence \\
GRB 100224A & a & likely missed due to very short/low-fluence prompt emission &
No record in GBM catalogs/GCNs &
Likely via \swift/BAT refined analysis (GCN) &
No evidence \\
GRB 101129A & a & likely missed due to very short/low-fluence prompt emission &
Yes – triggered (GBM) &
Yes ({\it Fermi}/GBM catalog \& GCN) &
Yes \\
GRB 120817B & a & likely missed due to very short/low-fluence prompt emission &
Yes – triggered (GBM) &
Yes ({\it Fermi}/GBM catalog \& GCN) &
Yes \\
GRB 140402A & a & likely missed due to very short/low-fluence prompt emission &
Yes – triggered (GBM/LAT) &
Yes ({\it Fermi}/GBM catalog \& GCN) &
No evidence \\
GRB 180718A & a & likely missed due to very short/low-fluence prompt emission &
Yes – triggered (GBM) &
Yes ({\it Fermi}/GBM catalog \& GCN) &
Yes \\
GRB 050906 & b & probably not a GRB (e.g., SGR/other) &
No – pre-\textit{Fermi} &
Likely via \swift/BAT refined analysis (GCN) &
No evidence \\
GRB 050925 & b & probably not a GRB (e.g., SGR/other) &
No – pre-\textit{Fermi} &
Likely via \swift/BAT refined analysis (GCN) &
No evidence \\
GRB 051105A & b & probably not a GRB (e.g., SGR/other) &
No – pre-\textit{Fermi} &
Likely via \swift/BAT refined analysis (GCN) &
No evidence \\
GRB 070209 & b & probably not a GRB (e.g., SGR/other) &
No – pre-\textit{Fermi} &
Likely via \swift/BAT refined analysis (GCN) &
No evidence \\
GRB 070810B & b & probably not a GRB (e.g., SGR/other) &
No – pre-\textit{Fermi} &
Likely via \swift/BAT refined analysis (GCN) &
No evidence \\
GRB 100628A & b & probably not a GRB (e.g., SGR/other) &
No record in GBM catalogs/GCNs &
Likely via \swift/BAT refined analysis (GCN) &
No evidence \\
GRB 130626A & b & probably not a GRB (e.g., SGR/other) &
Yes – triggered (GBM) &
Yes ({\it Fermi}/GBM catalog \& GCN) &
No evidence \\
GRB 170112A & b & probably not a GRB (e.g., SGR/other) &
No record in GBM catalogs/GCNs &
Likely via \swift/BAT refined analysis (GCN) &
No evidence \\
GRB 050202 & c & incomplete follow-up (constraints) &
No – pre-\textit{Fermi} &
Likely via \swift/BAT refined analysis (GCN) &
No evidence \\
GRB 070923 & c & incomplete follow-up (constraints) &
No – pre-\textit{Fermi} &
Likely via \swift/BAT refined analysis (GCN) &
No evidence \\
GRB 071112B & c & incomplete follow-up (constraints) &
No – pre-\textit{Fermi} &
Likely via \swift/BAT refined analysis (GCN) &
No evidence \\
GRB 081101 & c & incomplete follow-up (constraints) &
Yes – triggered (GBM) &
Yes ({\it Fermi}/GBM catalog \& GCN) &
No evidence \\
GRB 090417A & c & incomplete follow-up (constraints) &
No record in GBM catalogs/GCNs &
Likely via \swift/BAT refined analysis (GCN) &
No evidence \\
GRB 110420B & c & incomplete follow-up (constraints) &
Yes – triggered (GBM) &
Yes ({\it Fermi}/GBM catalog \& GCN) &
No evidence \\
GRB 111126A & c & incomplete follow-up (constraints) &
No record in GBM catalogs/GCNs &
Likely via \swift/BAT refined analysis (GCN) &
No evidence \\
GRB 120229A & c & incomplete follow-up (constraints) &
No record in GBM catalogs/GCNs &
Likely via \swift/BAT refined analysis (GCN) &
No evidence \\
GRB 140414A & c & incomplete follow-up (constraints) &
Unknown/ambiguous &
Likely via \swift/BAT refined analysis (GCN) &
No evidence \\
GRB 140606A & c & incomplete follow-up (constraints) &
Yes – subthreshold/ground (GBM) &
Possible (GBM targeted/ground search; check CTTE products) &
No evidence \\
GRB 151228A & c & incomplete follow-up (constraints) &
Yes – triggered (GBM) &
Yes ({\it Fermi}/GBM catalog \& GCN) &
No evidence \\
GRB 160612A & c & incomplete follow-up (constraints) &
Yes – triggered (GBM) &
Yes ({\it Fermi}/GBM catalog \& GCN) &
Yes \\
GRB 160726A & c & incomplete follow-up (constraints) &
Yes – triggered (GBM) &
Yes ({\it Fermi}/GBM catalog \& GCN) &
Yes \\
GRB 170325A & c & incomplete follow-up (constraints) &
Yes – triggered (GBM) &
Yes ({\it Fermi}/GBM catalog \& GCN) &
No evidence \\
GRB 180715A & c & incomplete follow-up (constraints) &
Yes – triggered (GBM) &
Yes ({\it Fermi}/GBM catalog \& GCN) &
No evidence \\
GRB 081226B & d & likely instrumental/data-quality issue &
Yes – GBM observation &
Yes ({\it Fermi}/GBM catalog \& GCN) &
No evidence \\
GRB 110112B & d & likely instrumental/data-quality issue &
Yes – triggered (GBM) &
Yes ({\it Fermi}/GBM catalog \& GCN) &
No evidence \\
GRB 131224A & d & likely instrumental/data-quality issue &
No record in GBM catalogs/GCNs &
Likely via \swift/BAT refined analysis (GCN) &
No evidence \\
\bottomrule
\end{tabular}
\end{sidewaystable*}

\section{Methodology}
\label{sec:method}

To test the hypothesis that an observed transient originates from the terminal evaporation of a primordial black hole (PBH), we developed a forward-modeling framework that enables a direct, quantitative comparison between theoretical predictions and observed gamma-ray burst (GRB) light curves.  
This approach combines physical modeling of the PBH photon emission, realistic detector-response folding, background estimation, and statistical model comparison against standard empirical GRB pulse templates.

\subsection{Theoretical photon counts}
\label{sec:theory}

A non-rotating, uncharged black hole of mass $M$ radiates thermally with a characteristic Hawking temperature,
\begin{equation}
T_H=\frac{1}{4\pi G_N M}\simeq1.06
\left(\frac{10^6\,{\rm g}}{M}\right)\,{\rm GeV},
\end{equation}
so that as the mass decreases the temperature and luminosity rise rapidly, leading to runaway evaporation.  
The differential emission rate for species $i$ of spin $s_i$ is
\begin{equation}
\frac{\partial^2N_i}{\partial E_i\partial t}
   =\frac{1}{2\pi}
   \frac{\Gamma_i(E_i,M)}{\exp(E_i/T_H)-(-1)^{2s_i}},
\end{equation}
where $\Gamma_i(E_i,M)$ is the grey-body factor accounting for frequency-dependent transmission through the black-hole potential barrier~\cite{Page1976}.  
The dominant contribution at late times comes from photons, electrons, and quarks that hadronize into $\pi^0$ mesons whose decay feeds the $\gamma$-ray spectrum~\cite{MacGibbonWebber1990,Heckler1997}.  

We generate the instantaneous photon and secondary spectra using the public code \texttt{BlackHawk}~\cite{Arbey:2019cmf}, which implements the full Standard-Model particle content and interfaces with \texttt{PYTHIA}~\cite{Sjostrand:2014zea} to simulate fragmentation and decays.  
Integrating over the time-evolving mass $M(t)$ yields the photon flux as a function of energy and remaining lifetime.  
As noted above, because $\dot M\propto -M^{-2}$, the emitted power $L\propto M^{-2}$ diverges toward the end of evaporation, producing a nearly universal luminosity profile
\begin{equation}
C_\gamma(t_{\rm rem})\propto t_{\rm rem}^{-p},\qquad p\simeq0.3\text{--}0.4,
\end{equation}
where $t_{\rm rem}$ is the remaining time to evaporation and the exponent $p$ depends only weakly on energy band and particle content~\cite{MacGibbonCarr1991}.  
This universality allows the use of a single normalized PBH template—scaled by total fluence and onset time—for all sub-second events in the relevant mass range.

We denote by $C^{\rm th}_{j,i}(\boldsymbol{\theta})$ the theoretical photon counts in energy bin $j$ and time bin $i$, parameterized by the vector $\boldsymbol{\theta}$ (normalization, onset time, and any nuisance parameters).  
For the nearby events of interest ($d\!\lesssim\!1$ pc), redshift and absorption corrections are negligible.

\subsection{Background modeling}
\label{sec:background}

Accurate background estimation is essential because PBH-like bursts are expected to be faint and short.  
For each detector we estimate the background level from pre- and post-burst intervals well outside the putative signal window.  
These counts are fit with a low-order polynomial or exponential model $b_{d,c,i}(t)$ that tracks orbital and detector trends.  
Uncertainties in the background fit are propagated by introducing Gaussian priors on $b_{d,c,i}$ or by marginalizing over nuisance parameters in the likelihood.  
This treatment, combined with Poisson statistics where appropriate~\cite{Cash:1979vz}, provides an unbiased measure of significance even in low-count regimes.

\subsection{Instrumental response and forward folding}
\label{sec:response}

To compare the predicted photon flux with the observed counts, we fold the theoretical spectra through each instrument’s response matrix.  
For detector $d$ and energy channel $c$ the expected counts are
\begin{equation}
\mu_{d,c,i}(\boldsymbol{\theta})
 = \sum_{j}K^{(i)}_{d,c\leftarrow j}
   \,C^{\rm th}_{j,i}(\boldsymbol{\theta})
   + b_{d,c,i},
\end{equation}
where $K^{(i)}_{d,c\leftarrow j}$ encodes effective area, energy redistribution, and temporal exposure.  
For \swift/BAT we use the official HEASARC response matrices; for \gbm, we employ time-dependent responses produced with \texttt{GBMRSP}.  
This “forward-folding’’ approach ensures that comparisons with data are made in detector space rather than in photon space, preserving instrumental resolution and selection effects.

\subsection{Likelihood analysis}
\label{sec:likelihood}

Observed counts $n_{d,c,i}$ are compared to $\mu_{d,c,i}(\boldsymbol{\theta})$ via a Gaussian or Poisson likelihood,
\begin{equation}
\ln\mathcal{L}(\boldsymbol{\theta})
 = -\tfrac12
   \sum_{d,c,i}
   \frac{[n_{d,c,i}-\mu_{d,c,i}(\boldsymbol{\theta})]^2}
        {\sigma_{d,c,i}^2}
   + \mathrm{const.},
\end{equation}
or, for sparse bins,
\begin{equation}
\ln\mathcal{L}_\mathrm{C}
 = \sum_{d,c,i}\!\left[
   n_{d,c,i}\ln\mu_{d,c,i}
   -\mu_{d,c,i}
   -\ln(n_{d,c,i}!)
   \right].
\end{equation}
We maximize $\mathcal{L}$ or sample it using Markov-chain Monte Carlo to infer posterior distributions of $\boldsymbol{\theta}$ and to compute evidence ratios between competing models.  
This framework allows us to test whether the data favor the PBH “backwards-burst’’ morphology relative to empirical GRB pulse shapes.

\subsection{Model comparison and validation}
\label{sec:modelcomp}

Each burst is fit with four models:  
(i) the physically motivated PBH template described above,  
(ii) a Fast-Rise–Exponential-Decay (FRED) pulse~\cite{Kocevski2003},  
(iii) an Exponential-Rise–Constant-Afterglow (ERCA) form~\cite{RydeSvensson2002}, and
(iiii) a PBH-Constant-Afterglow (PBHCA) model.  
For the FRED model, rise and decay constants $(\tau_1,\tau_2)$ and asymmetry $\kappa=\tau_1/(\tau_1+\tau_2)$ describe the temporal profile.  
The ERCA model,
\begin{equation}
I(t)=A\,\exp\!\left[-\frac{\tau_1}{t-t_s}\right],
\end{equation}
captures events with extended emission or quasi-steady tails.  
Model preference is quantified using the Akaike (AIC) and Bayesian (BIC) information criteria:
\begin{align}
{\rm AIC}_M &= \chi_M^2 + 2k_M,\\
{\rm BIC}_M &= \chi_M^2 + k_M\ln n,
\end{align}
where $k_M$ is the number of free parameters and $n$ the number of bins.  
We validate the analysis pipeline by injecting synthetic PBH light curves into real background segments; recovered parameters confirm sensitivity to sub-second, hard, and monotonic pulses.

Together, these steps define a physically grounded and statistically rigorous framework for identifying or excluding PBH-evaporation candidates in short-GRB catalogs.

\section{Results}
\label{sec:results}

We applied the forward-folded PBH evaporation model and the two empirical GRB templates (FRED and ERCA) to the sample of 35 short \swift\ bursts lacking detected afterglows (see Table~\ref{tab:list}).  
For each event, we optimized the model parameters using the maximum-likelihood procedure outlined in Sec.~\ref{sec:likelihood} and evaluated the relative quality of fit via the AIC and BIC statistics.  
The resulting best-fit parameters and model-selection outcomes are summarized in Table~\ref{tab:fitresults}.

\subsection{Light-curve morphology and characteristic timescales}

As mentioned above, the general theoretical expectation for the terminal PBH evaporation lightcurve features a steep monotonic brightening as the remaining time to evaporation approaches zero, followed by an abrupt cutoff.  
When compared with the observed \swift/BAT light curves, virtually all bursts in the analyzed sample exhibit the opposite trend—rapid rise and slower decay—characteristic of conventional GRB pulses.

Table~\ref{tab:fitresults} quantifies this behavior through the ratio $\tau_{\mathrm{rise}}/\tau_{\mathrm{dec}}$.  
Every event in the sample shows $\tau_{\mathrm{rise}}/\tau_{\mathrm{dec}}<1$, typically in the range $0.1$–$0.5$, corresponding to asymmetric, fast-rise–slow-decay profiles.  
Such temporal development is consistent with the FRED template and inconsistent with the PBH “backwards-burst’’ expectation, which predicts a slow-rise–fast-decay evolution when expressed in observer time.

The characteristic rise and decay times span $\sim0.01$–$0.5$ s, implying that the emission episodes are short but resolvable by the \swift/BAT 64 ms binned light curves.  
These durations are typical of the short-GRB population and do not suggest an additional, distinct ultra-short sub-class that would be expected for nearby PBH explosions ($T_{90}\!\lesssim\!100$ ms).

\subsection{Model-comparison statistics}

The AIC and BIC values reported in Table~\ref{tab:fitresults} provide quantitative evidence against the PBH hypothesis for this dataset.  
For all 35 bursts, both criteria favor either the FRED or ERCA models over the PBH template.  
The differences are substantial: typical $\Delta{\rm AIC}$ and $\Delta{\rm BIC}$ values exceed $10^2$–$10^3$, far above the conventional threshold ($\Delta>10$) for strong preference.  
In no case does the PBH template yield a lower information criterion than either empirical model.  

The two phenomenological descriptions (FRED and ERCA) together capture the diversity of observed pulse shapes.  
Approximately two-thirds of the events are best fit by a FRED profile, while the remainder, generally those with flatter post-peak behavior, are marginally better described by ERCA.  
In both subsets, the reduced $\chi^2$ values cluster around unity, indicating statistically adequate fits.  
In contrast, PBH template fits systematically underpredict the pre-peak flux and overpredict the decay phase, producing large residuals and elevated $\chi^2$ values ($\gtrsim10^3$ in many cases).

\subsection{Representative examples}

Figure~\ref{fig:fitexamples} shows four representative short bursts fitted with the three models.  
The FRED and ERCA curves (orange and green) reproduce the observed asymmetric pulses and gradual decays with high fidelity, whereas the PBH template (blue) yields a poor match, missing both the temporal asymmetry and the observed curvature in the peak region.  
These residual patterns are typical across the sample: the PBH light curve lacks sufficient flexibility to mimic the observed structure, which in several cases includes secondary peaks or extended tails inconsistent with a monotonic rise.

The lower panels of Fig.~\ref{fig:fitexamples} illustrate the corresponding residuals, reinforcing the visual conclusion that the PBH model cannot reproduce the observed temporal evolution even when normalization and timing are optimized.  
If the PBH evaporation scenario were relevant to any of these events, one would expect at least a few cases where the PBH template provided a statistically comparable or superior fit, but none are found.

\begin{figure*}
\centering
\includegraphics[width=0.85\textwidth]{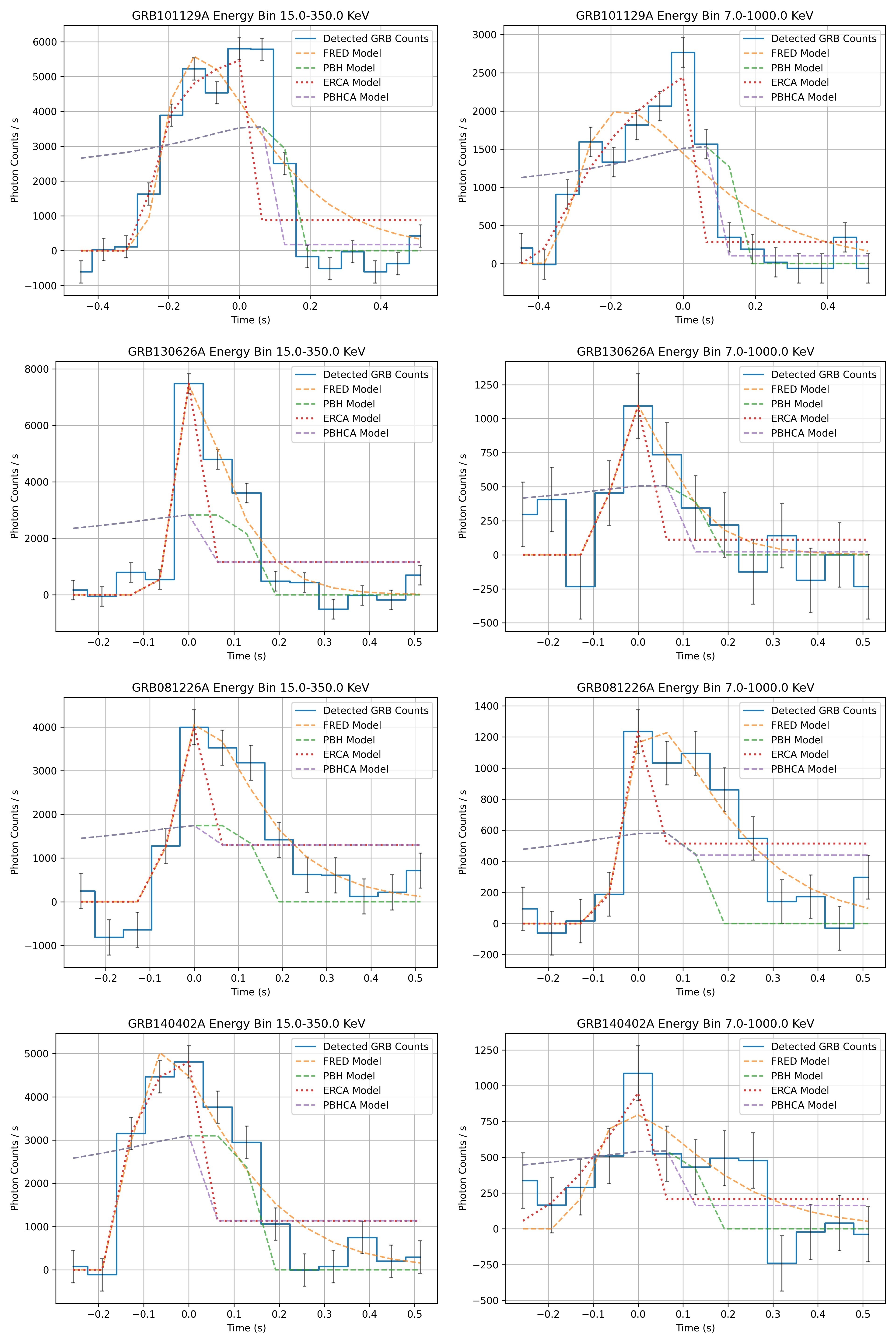}
\caption{\textbf{Comparison of light curves for the GRB events in Table 1 observed by \textit{Swift}/BAT and \textit{Fermi}/GBM. }Four GRB events detected by both \swift/BAT and \textit{Fermi}/GBM at different energy bin are compared and fitted with four models. }
\label{fig:swiftvsfermi}
\end{figure*}

\begin{figure*}[t]
\centering
\includegraphics[width=0.85\textwidth]{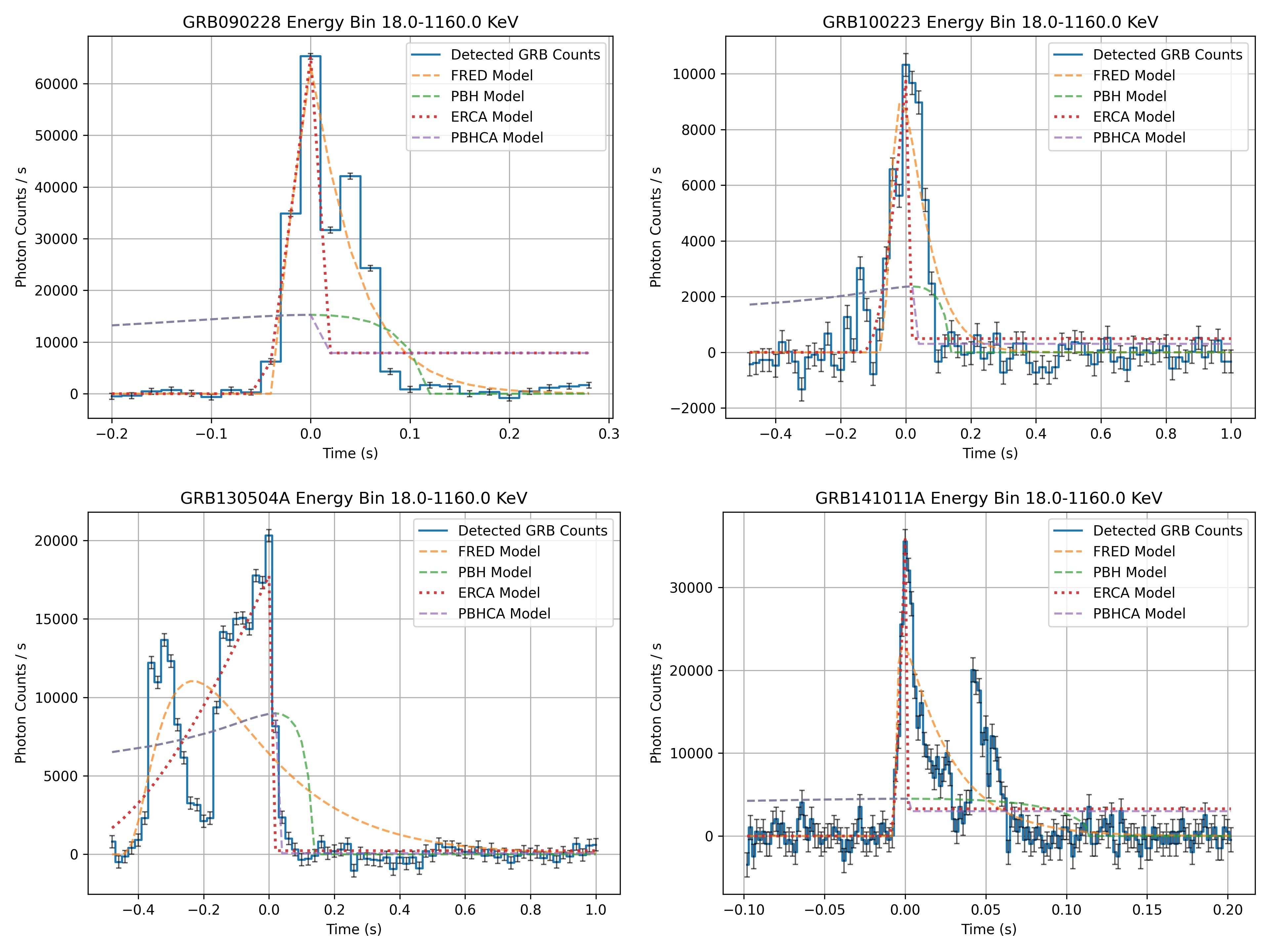}
\caption{\textbf{Analysis of GRB events potentially close to Earth ({\it Fermi} data)}. Four GRB events identified by the IPN as compatible with a location close to Earth are modeled. The estimated minimum distance from Earth for these events is less than $10^{14}$cm.}
\label{fig:grbcloseby}
\end{figure*}

Figure~\ref{fig:swiftvsfermi} presents the light curves of four GRB events observed in both the \swift and {\it Fermi} catalogs. We also analyzed the {\it Fermi} FITS files for four additional GRB events detected with IPN, whose curves are shown in Figure~\ref{fig:grbcloseby}. IPN triangulation for these events indicates that they may have possibly occurred at a distance from Earth compatible with the luminosity of a PBH explosion event. Although the \swift and {\it Fermi} detectors operate with different energy bins, we arrived at conclusions consistent with those discussed above.

\subsection{Implications for PBH–evaporation searches}
\label{subsec:rate_limit}

The absence of any burst whose temporal morphology is even marginally
compatible with the PBH ``backwards–burst'' template in our sample of
35 \swift short GRBs without detected afterglows can be translated into
an upper bound on the local rate density of terminal PBH evaporation
events.  The argument follows the standard Poisson–statistics treatment
used in previous searches for evaporating PBHs with VHE $\gamma$–ray
observatories and the IPN.

We parameterize the local explosion rate density as
\begin{equation}
  R_{\rm PBH} \equiv \frac{{\rm d}N_{\rm exp}}{{\rm d}V\,{\rm d}t}
  \quad [{\rm pc}^{-3}\,{\rm yr}^{-1}]\,.
\end{equation}
For an instrument with an effective sky coverage fraction $f_{\rm sky}$,
an observing time $T_{\rm obs}$, and a maximum distance $d_{\max}$ out
to which a PBH explosion can trigger the detector and pass our analysis
cuts, the expected number of detectable PBH events is
\begin{equation}
  \lambda \equiv N_{\rm exp}
  \simeq R_{\rm PBH}\;
           V_{\rm eff}\;
           T_{\rm eff}
  = R_{\rm PBH}\;
    \left(\frac{4\pi}{3} d_{\max}^3\right)
    \left(f_{\rm sky}\,T_{\rm obs}\right)\!,
  \label{eq:lambda_def}
\end{equation}
where we have assumed a homogeneous local PBH distribution and
isotropic detectability within $d_{\max}$.

In the absence of any PBH–like candidates, the observed number of
events is $N_{\rm obs}=0$, so the Poisson likelihood
$P(N_{\rm obs}|\lambda)=\exp(-\lambda)$ implies
$\lambda \lesssim \lambda_{95}\simeq 3$ at 95\% confidence.  Solving
Eq.~\eqref{eq:lambda_def} for $R_{\rm PBH}$ yields the general
constraint
\begin{equation}
  R_{\rm PBH} \lesssim
  \frac{\lambda_{95}}{(4\pi/3)\,d_{\max}^3\,f_{\rm sky}\,T_{\rm obs}}\,.
  \label{eq:R_general}
\end{equation}

To estimate $d_{\max}$ we compare the expected fluence of a terminal
PBH evaporation event with the typical {\it Swift}/BAT fluence threshold for
short GRBs.  The total energy emitted in $\gamma$ rays in the BAT band
for a standard–model PBH approaching its end point is
\begin{equation}
  E_{\gamma}^{\rm BAT} \sim 10^{28}\text{--}10^{29}\ {\rm erg}\,,
\end{equation}
as obtained from the time–integrated {\tt BlackHawk} spectra including
secondary hadronic cascades and folded through the BAT response
(see Sec.~\ref{sec:theory}).  For a limiting fluence
$F_{\rm lim}\sim 10^{-7}\ {\rm erg\ cm^{-2}}$ characteristic of the
faintest short \swift bursts in our sample, the corresponding horizon
distance is
\begin{eqnarray}
  \nonumber d_{\max} &\simeq&
  \left[\frac{E_{\gamma}^{\rm BAT}}{4\pi F_{\rm lim}}\right]^{1/2}
  \simeq 0.01\ {\rm pc}\;\\
  &&\left(\frac{E_{\gamma}^{\rm BAT}}{10^{28}\ {\rm erg}}\right)^{1/2}
  \left(\frac{F_{\rm lim}}{10^{-7}\ {\rm erg\ cm^{-2}}}\right)^{-1/2}\!.
  \label{eq:dmax_scaling}
\end{eqnarray}
This estimate is consistent with previous horizon calculations for
exploding PBHs in the keV–MeV band. 

{\it Swift}/BAT has monitored the sky for a time
$T_{\rm obs}\simeq 14\ {\rm yr}$ over the interval considered here.
The BAT partially coded field of view and duty cycle correspond to an
effective sky coverage fraction $f_{\rm sky}\sim 0.1$ for short,
hard transients that satisfy our selection cuts. Inserting the
fiducial values $d_{\max}\simeq 10^{-2}\ {\rm pc}$,
$f_{\rm sky}\simeq 0.1$, $T_{\rm obs}\simeq 14\ {\rm yr}$ and
$\lambda_{95}\simeq 3$ into Eq.~\eqref{eq:R_general} gives
\begin{eqnarray}
   \nonumber R_{\rm PBH}
  &\lesssim
  \frac{3}
       {(4\pi/3)\,(10^{-2}\ {\rm pc})^3\,
        (0.1)\,(14\ {\rm yr})}
  \\[3pt]
  &\simeq {\cal O}\!\left(10^{5}\right)\ {\rm pc^{-3}\ yr^{-1}}\,.
  \label{eq:R_numeric}
\end{eqnarray}
Retaining explicit parameter dependences, the bound can be written as
\begin{eqnarray}
  \nonumber R_{\rm PBH} &\lesssim&
  1\times 10^{5}\ {\rm pc^{-3}\ yr^{-1}}\,
  \left(\frac{\lambda_{95}}{3}\right)\times\\
  &&\left(\frac{f_{\rm sky}}{0.1}\right)^{-1}
  \left(\frac{T_{\rm obs}}{14\ {\rm yr}}\right)^{-1}
  \left(\frac{d_{\max}}{10^{-2}\ {\rm pc}}\right)^{-3}\!,
  \label{eq:R_scaling}
\end{eqnarray}
which makes clear that our constraint is an order–of–magnitude
statement, limited chiefly by the uncertainty in the effective
detection volume for PBH–like events.

The bound in Eq.~\eqref{eq:R_numeric} is comparable, within factors of a few, to the limits obtained
from dedicated searches for PBH bursts with {\it Fermi}/GBM and
TeV $\gamma$–ray instruments such as Milagro, HAWC, and H.E.S.S.
\cite{Abdo:2015osa,Albert:2019lkt,HESS:2021pbh,Carr:2016hva}.
Given the simplifying assumptions about the PBH emission spectrum and
the \swift triggering efficiency for ultra–short, hard events, we
consider Eq.~\eqref{eq:R_numeric} a conservative, instrument–specific
constraint on the local PBH evaporation rate density.

\subsection{Summary of findings}

In summary:
\begin{itemize}
    \item All 35 analyzed short GRBs without afterglows display temporal asymmetries $\tau_{\mathrm{rise}}/\tau_{\mathrm{dec}}<1$, inconsistent with the slow-rise PBH signature.
    \item The PBH “backwards-burst’’ template fails to reproduce the observed pulse morphology, yielding systematically poorer likelihood and information-criterion values than FRED or ERCA fits.
    \item No event exhibits the monotonic brightening or spectral hardening expected for terminal PBH evaporation.
\end{itemize}
We therefore identify no credible PBH-evaporation candidates in this dataset.  
The analysis framework developed here nonetheless establishes a template-based methodology that can be extended to larger samples and higher-energy instruments (e.g., \gbm, HAWC, or CTA) to further constrain the local PBH explosion rate.

\begin{figure*}[t]
\centering
\includegraphics[width=0.85\textwidth]{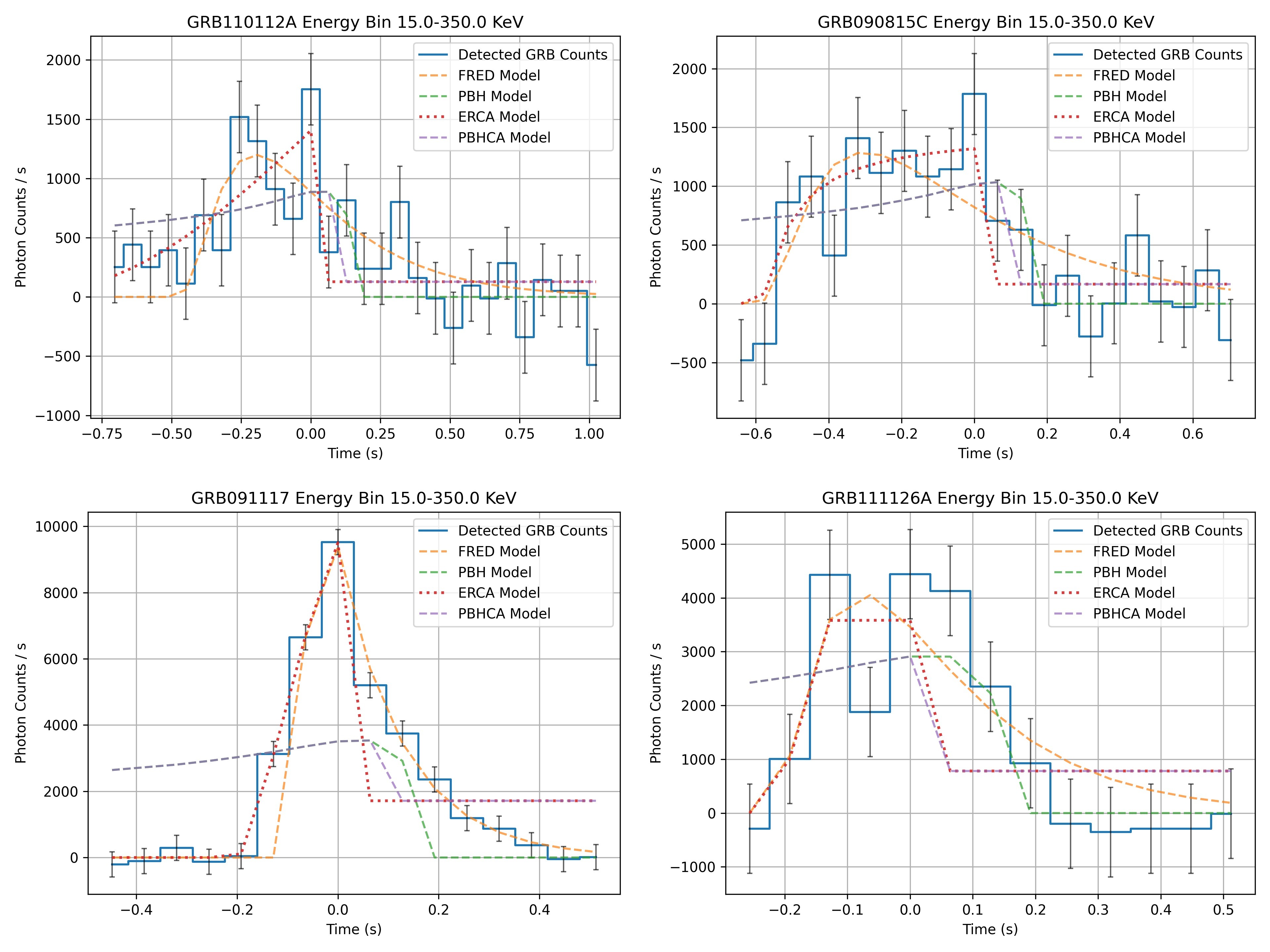}
\caption{
\textbf{Representative short-duration GRB light-curve fits.}  
Observed \swift/BAT count rates (solid blue) are fitted with four models: the physically motivated PBH evaporation template (dashed green), a Fast-Rise–Exponential-Decay (FRED) profile (dashed orange), an Exponential-Rise–Constant-Afterglow (ERCA) form (dashed red), and a PBH with Constant-Afterglow (PBHCA) curve (dashed purple).  
For all four examples the FRED or ERCA models reproduce the observed pulse asymmetry and extended decay far better than the PBH template, whose universal monotonic brightening shape cannot mimic the data.  
Residuals (bottom panels) highlight the systematic deviation of the PBH model, particularly its underprediction of the rising phase and overprediction of the decay tail.
}
\label{fig:fitexamples}
\end{figure*}

\begin{sidewaystable*}
\centering
\scriptsize
\setlength{\tabcolsep}{3pt}
\renewcommand{\arraystretch}{1.15}
\caption{sGRB light-curve analysis summary. For each GRB we report the best-fit parameters of the FRED and ERCA models, the AIC and BIC values for all four models (FRED, PBH, PBHCA, ERCA), and the preferred model according to AIC and BIC.}
\label{tab:fitresults}
\begin{tabular}{@{}l
                c c c c c c c
                c c
                c c
                c c c c
                c c@{}}
\toprule
\multirow{2}{*}{GRB event}
 & \multicolumn{7}{c}{FRED model}
 & \multicolumn{2}{c}{PBH model}
 & \multicolumn{2}{c}{PBHCA model}
 & \multicolumn{4}{c}{ERCA}
 & \multicolumn{2}{c}{Preferred model} \\
\cmidrule(lr){2-8} \cmidrule(lr){9-10} \cmidrule(lr){11-12} \cmidrule(lr){13-16} \cmidrule(lr){17-18}
 & $\tau_1$ (s) & $\tau_2$ (s) & $\tau_{\rm rise}$ (s) & $\tau_{\rm dec}$ (s) & $\tau_{\rm rise}/\tau_{\rm dec}$ & AIC & BIC
 & AIC & BIC
 & AIC & BIC
 & $\tau_1$ (s) & Const. & AIC & BIC
 & by AIC & by BIC \\
\midrule
GRB110112A  & 0.453 & 0.252 & 0.1918 & 0.4438 & 0.43 &  53 &  59 &   65 &   66 &   68 &   70 & 5.287 & 24.31 &  53 &  57 & FRED & ERCA \\
GRB090815C  & 0.546 & 0.310 & 0.2343 & 0.5443 & 0.43 &  52 &  58 &   69 &   70 &   74 &   76 & 0.208 & 63.00 &  42 &  47 & ERCA & ERCA \\
GRB091117   & 0.180 & 0.120 & 0.0857 & 0.2057 & 0.42 &  70 &  75 & 1009 & 1010 & 1095 & 1097 & 0.634 & 26.62 & 264 & 268 & FRED & FRED \\
GRB111126A  & 0.216 & 0.152 & 0.1065 & 0.2585 & 0.41 & 242 & 248 &  278 &  279 &  373 &  374 & 0.241 &  1.00 & 340 & 344 & FRED & FRED \\
GRB101129A  & 0.232 & 0.170 & 0.1175 & 0.2875 & 0.41 & 194 & 199 &  678 &  679 &  885 &  887 & 4.374 &  9.31 & 349 & 353 & FRED & FRED \\
GRB130626A  & 0.098 & 0.074 & 0.0506 & 0.1246 & 0.41 &  51 &  57 &  615 &  616 &  285 &  289 & 0.145 & 37.59 & 285 & 290 & FRED & FRED \\
GRB140414A  & 0.134 & 0.103 & 0.0700 & 0.1730 & 0.40 & 384 & 390 & 1039 & 1041 & 1288 & 1289 & 0.135 & 39.31 & 641 & 645 & FRED & FRED \\
GRB080121   & 0.148 & 0.114 & 0.0774 & 0.1914 & 0.40 &  58 &  63 &  103 &  105 &  130 &  132 & 1.988 & 17.50 &  81 &  86 & FRED & FRED \\
GRB081226A  & 0.149 & 0.115 & 0.0780 & 0.1930 & 0.40 &  71 &  77 &  397 &  399 &  386 &  387 & 0.263 & 21.47 & 214 & 219 & FRED & FRED \\
GRB140402A  & 0.168 & 0.132 & 0.0890 & 0.2210 & 0.40 &  76 &  82 &  552 &  554 &  624 &  626 & 0.759 & 74.22 & 222 & 226 & FRED & FRED \\
GRB081101   & 0.117 & 0.094 & 0.0628 & 0.1568 & 0.40 &  38 &  44 &  636 &  638 &  377 &  381 & 0.206 & 52.34 & 377 & 381 & FRED & FRED \\
GRB140606A  & 0.162 & 0.132 & 0.0878 & 0.2198 & 0.40 &  67 &  73 &  332 &  333 &  355 &  357 & 0.286 &  9.72 & 246 & 250 & FRED & FRED \\
GRB160726A  & 0.107 & 0.088 & 0.0583 & 0.1463 & 0.40 & 874 & 880 & 3537 & 3538 & 3841 & 3842 & 0.224 & 34.81 &1054 &1058 & FRED & FRED \\
GRB160612A  & 0.080 & 0.067 & 0.0441 & 0.1111 & 0.40 & 238 & 244 & 3570 & 3571 & 4148 & 4149 & 0.464 &  1.81 &1116 &1121 & FRED & FRED \\
GRB180718A  & 0.031 & 0.039 & 0.0222 & 0.0612 & 0.36 &  43 &  48 &  202 &  203 &  638 &  640 & 0.179 & 18.03 &  52 &  56 & FRED & FRED \\
GRB151228A  & 0.071 & 0.142 & 0.0679 & 0.2099 & 0.32 &  95 & 101 &  956 &  958 & 1005 & 1006 & 0.321 & 68.28 & 533 & 537 & FRED & FRED \\
GRB180715A  & 0.003 & 0.008 & 0.0034 & 0.0114 & 0.30 &  28 &  33 & 7286 & 7287 & 6628 & 6630 & 0.206 & -4.38 &  25 &  28 & ERCA & ERCA \\
GRB170112A  & 0.012 & 0.046 & 0.0171 & 0.0631 & 0.27 &  30 &  36 &  223 &  225 &  251 &  253 & 0.262 & 36.66 &  80 &  84 & FRED & FRED \\
GRB071112B  & 0.025 & 0.106 & 0.0379 & 0.1439 & 0.26 & 180 & 185 &  550 &  552 &  652 &  653 & 0.144 & 27.75 & 357 & 361 & FRED & FRED \\
GRB050906   & 0.001 & 0.007 & 0.0020 & 0.0090 & 0.23 &  45 &  51 &   58 &   59 &   58 &   59 & 0.197 &  1.59 &  43 &  47 & ERCA & ERCA \\
GRB131224B  & 0.001 & 0.017 & 0.0034 & 0.0204 & 0.17 &  36 &  42 &  466 &  467 &  480 &  482 & 0.311 & 60.03 &  60 &  64 & FRED & FRED \\
GRB051105A  & 0.001 & 0.023 & 0.0041 & 0.0271 & 0.15 &  35 &  41 &  205 &  206 &  217 &  219 & 0.175 & 12.56 &  82 &  87 & FRED & FRED \\
GRB100628A  & 0.001 & 0.024 & 0.0042 & 0.0282 & 0.15 &  58 &  64 &  712 &  713 &  703 &  704 & 0.047 & 24.66 &  48 &  52 & ERCA & ERCA \\
GRB120817B  & 0.001 & 0.028 & 0.0046 & 0.0326 & 0.14 & 104 & 108 & 14136&14137& 9772& 9773 & 1.860 & 39.59 & 166 & 168 & FRED & FRED \\
GRB050925   & 0.001 & 0.033 & 0.0050 & 0.0380 & 0.13 &  35 &  41 &  743 &  744 &  749 &  751 & 0.098 &  8.16 &  42 &  47 & FRED & FRED \\
GRB110420B  & 0.001 & 0.038 & 0.0054 & 0.0434 & 0.12 &  46 &  52 &  608 &  609 &  160 &  164 & 0.190 & 27.06 & 160 & 165 & FRED & FRED \\
GRB070810B  & 0.001 & 0.046 & 0.0060 & 0.0520 & 0.12 &  46 &  52 &  261 &  262 &  275 &  276 & 0.219 &  6.72 &  57 &  62 & FRED & FRED \\
GRB070923   & 0.001 & 0.047 & 0.0061 & 0.0531 & 0.11 &  55 &  61 & 1023 & 1025 & 1049 & 1050 & 0.248 & 41.09 &  97 & 101 & FRED & FRED \\
GRB070209   & 0.001 & 0.051 & 0.0064 & 0.0574 & 0.11 &  31 &  37 &  217 &  218 &  236 &  237 & 0.224 & 17.19 &  79 &  83 & FRED & FRED \\
GRB090417A  & 0.001 & 0.053 & 0.0065 & 0.0595 & 0.11 &  54 &  60 &  409 &  411 &  449 &  451 & 0.071 &  9.38 & 172 & 177 & FRED & FRED \\
GRB100216A  & 0.001 & 0.075 & 0.0078 & 0.0828 & 0.09 &  34 &  40 &  286 &  287 &  300 &  301 & 0.183 & 54.81 &  61 &  66 & FRED & FRED \\
GRB050202   & 0.001 & 0.094 & 0.0089 & 0.1029& 0.09 &  32 &  38 &  190 &  192 &  209 &  211 & 0.321 & 27.16 &  74 &  78 & FRED & FRED \\
GRB120229A  & 0.001 & 0.171 & 0.0122 & 0.1832& 0.07 & 172 & 177 &  487 &  489 &  487 &  489 & 0.019 & 44.41 & 241 & 245 & FRED & FRED \\
GRB170325A  & 0.001 & 0.175 & 0.0124 & 0.1874& 0.07 &  73 &  79 & 1304 &1305 & 1283 &1285 & 0.235 & 61.03 & 613 & 618 & FRED & FRED \\
GRB100224A  & 0.001 & 0.256 & 0.0151 & 0.2711& 0.06 & 113 & 117 &  451 &  452 &  464 &  465 & 0.231 & 18.12 & 268 & 271 & FRED & FRED \\
\bottomrule
\end{tabular}
\end{sidewaystable*}

\section{Conclusions}
\label{sec:conclusions}

We have developed and applied a physically grounded, forward-modeling framework to search for signatures of terminal primordial black-hole (PBH) evaporation in short-duration gamma-ray bursts (sGRBs).  
Our approach directly confronts the universal temporal and spectral behavior predicted by Hawking radiation with real GRB light curves, accounting for instrumental response, background uncertainties, and statistical model comparison.

By simulating the full photon emission history of an evaporating PBH using the \texttt{BlackHawk} package, we derived the expected temporal profile of the photon count rate.  
The resulting ``backwards-burst'' morphology—monotonic flux increase culminating in a final cutoff—is a distinctive and essentially universal signature of the PBH evaporation process, insensitive to details such as initial mass, spin, or composition.  
This prediction provided a physically motivated template against which we tested short GRBs observed by \swift.

We analyzed a well-defined sample of 35 \swift\ short GRBs with no detected afterglows, drawn from the Dichiara et~al.~\cite{Dichiara2020} catalog and augmented through cross-correlation with {\it Fermi}/GBM, INTEGRAL, Konus-Wind, and the Interplanetary Network archives.  
This selection specifically targets events whose lack of afterglow or host association makes them plausible candidates for exotic progenitors, while ensuring adequate signal-to-noise for temporal-profile fitting.

The results of our model comparison are unambiguous.  
All bursts exhibit asymmetric, fast-rise–slow-decay light curves ($\tau_{\mathrm{rise}}/\tau_{\mathrm{dec}} < 1$), well described by standard empirical FRED or ERCA profiles but inconsistent with the slow-rise–fast-decay behavior predicted for an evaporating PBH.  
Across the sample, both the Akaike and Bayesian information criteria decisively favor the FRED or ERCA fits, with typical $\Delta$AIC and $\Delta$BIC values exceeding $10^2$–$10^3$ relative to the PBH template (see Table~\ref{tab:fitresults}).  
No individual event shows even marginal statistical preference for the PBH hypothesis.  
Representative examples (Figure~\ref{fig:fitexamples}) illustrate these differences clearly: the PBH template systematically underestimates early-time emission and overestimates late-time flux, failing to reproduce the observed pulse asymmetry.

These findings imply that none of the analyzed short GRBs are compatible with being final-stage PBH explosions.  
The null result allows us to place an empirical upper bound on the local rate density of such events.  
Adopting the \swift/BAT exposure and sensitivity appropriate for our analysis window, the non-detection of PBH-like transients over $\sim14$ years implies $R_{\mathrm{PBH}}\lesssim10^5\ \mathrm{pc^{-3}\,yr^{-1}}$, comparable to limits derived from {\it Fermi}/GBM and IPN searches~\cite{Carr:2016drx,Albert:2019lkt,HESS:2021pbh}.  
Although this constraint remains orders of magnitude above the rate expected if PBHs make a significant contribution to the dark matter density, it robustly confirms that no nearby, high-fluence evaporation events have occurred within the reach of current GRB detectors.

The methodology established here can be readily extended.  
Future searches can exploit:
\begin{itemize}
    \item the expanded temporal coverage and energy range of \textit{Fermi}/GBM, HAWC, and CTA;
    \item joint temporal–spectral analyses incorporating the predicted hardening trend near the evaporation endpoint;
    \item population-level Bayesian inference to combine sub-threshold or marginal events across instruments; and
    \item cross-correlation with archival all-sky monitors for multi-detector coincidence tests.
\end{itemize}
Such multi-mission analyses could lower rate limits by several orders of magnitude and potentially reveal faint, nearby PBH events if they exist.

In conclusion, our work provides both a null detection and a methodological foundation.  
It demonstrates that PBH evaporation signatures can be robustly searched for within GRB catalogs through physically motivated template fitting, and it outlines a path toward exploiting future, higher-sensitivity instruments to probe deeper into the final moments of evaporating primordial black holes.

\begin{acknowledgments}
This work is partly supported by the U.S.\ Department of Energy grant number de-sc0010107 (SP).
\end{acknowledgments}

\bibliographystyle{apsrev4-2}
\bibliography{refs}

@article{Hawking1974,
  author = {Hawking, S. W.},
  title = {Black hole explosions?},
  journal = {Nature},
  year = {1974},
  volume = {248},
  pages = {30--31},
  doi = {10.1038/248030a0}
}

@article{Hawking1975,
  author = {Hawking, S. W.},
  title = {Particle Creation by Black Holes},
  journal = {Communications in Mathematical Physics},
  year = {1975},
  volume = {43},
  number = {3},
  pages = {199--220},
  doi = {10.1007/BF02345020}
}

@article{MacGibbonCarr1991,
  author = {MacGibbon, Jane H. and Carr, Bernard J.},
  title = {Cosmic Rays from Primordial Black Holes},
  journal = {The Astrophysical Journal},
  year = {1991},
  volume = {371},
  pages = {447--469},
  doi = {10.1086/169909}
}

@article{Kouveliotou1993,
  author = {Kouveliotou, C. and Meegan, C. A. and Fishman, G. J. and Bhat, N. P. and Briggs, M. S. and Koshut, T. M. and Paciesas, W. S. and Pendleton, G. N.},
  title = {Identification of Two Classes of Gamma-Ray Bursts},
  journal = {The Astrophysical Journal Letters},
  year = {1993},
  volume = {413},
  pages = {L101--L104},
  doi = {10.1086/186969}
}

@article{SariPiranNarayan1998,
  author = {Sari, R. and Piran, T. and Narayan, R.},
  title = {Spectra and Light Curves of Gamma-Ray Burst Afterglows},
  journal = {The Astrophysical Journal Letters},
  year = {1998},
  volume = {497},
  pages = {L17--L20},
  doi = {10.1086/311269}
}

@article{Nakar2007,
  author = {Nakar, Ehud},
  title = {Short-Hard Gamma-Ray Bursts},
  journal = {Physics Reports},
  year = {2007},
  volume = {442},
  number = {1-6},
  pages = {166--236},
  doi = {10.1016/j.physrep.2007.02.005}
}

@article{Berger2014,
  author = {Berger, Edo},
  title = {Short-Duration Gamma-Ray Bursts},
  journal = {Annual Review of Astronomy and Astrophysics},
  year = {2014},
  volume = {52},
  pages = {43--105},
  doi = {10.1146/annurev-astro-081913-035926}
}

@article{Kaneko2015,
    author = {Kaneko, Y. and Bostanc{\i}, Z. F. and G{\"o}{\u{g}}{\"u}{\c{s}}, E. and Lin, L.},
    title = "{Short Gamma-Ray Bursts with Extended Emission Observed with Swift/BAT and Fermi/GBM}",
    eprint = "1506.05899",
    archivePrefix = "arXiv",
    primaryClass = "astro-ph.HE",
    doi = "10.1093/mnras/stv1286",
    journal = "Mon. Not. Roy. Astron. Soc.",
    volume = "452",
    number = "1",
    pages = "824--837",
    year = "2015"
}

@article{Dichiara2020,
    author = "Dichiara, S. and Troja, E. and O'Connor, B. and Marshall, F. E. and Beniamini, P. and Cannizzo, J. K. and Lien, A. Y. and Sakamoto, T.",
    title = "{Short gamma-ray bursts within 200 Mpc}",
    eprint = "1912.08698",
    archivePrefix = "arXiv",
    primaryClass = "astro-ph.HE",
    doi = "10.1093/mnras/staa124",
    journal = "Mon. Not. Roy. Astron. Soc.",
    volume = "492",
    number = "4",
    pages = "5011--5022",
    year = "2020"
}

@article{Meegan2009,
  author = {Meegan, C. and others},
  title = {The Fermi Gamma-Ray Burst Monitor},
  journal = {The Astrophysical Journal},
  year = {2009},
  volume = {702},
  number = {1},
  pages = {791--804},
  doi = {10.1088/0004-637X/702/1/791}
}

@article{Atwood2009,
  author = {Atwood, W. B. and others},
  title = {The Large Area Telescope on the Fermi Gamma-Ray Space Telescope Mission},
  journal = {The Astrophysical Journal},
  year = {2009},
  volume = {697},
  number = {2},
  pages = {1071--1102},
  doi = {10.1088/0004-637X/697/2/1071}
}

@article{Aptekar1995,
  author = {Aptekar, R. L. and others},
  title = {Konus-W Gamma-Ray Burst Experiment for the GGS Wind Spacecraft},
  journal = {Space Science Reviews},
  year = {1995},
  volume = {71},
  number = {1-4},
  pages = {265--272},
  doi = {10.1007/BF00751332}
}

@article{Winkler2003,
  author = {Winkler, C. and others},
  title = {The INTEGRAL mission},
  journal = {Astronomy \& Astrophysics},
  year = {2003},
  volume = {411},
  pages = {L1--L6},
  doi = {10.1051/0004-6361:20031288}
}

@article{Hurley2013,
  author = {Hurley, K. and others},
  title = {The Interplanetary Network Supplement to the Fermi GBM Catalog of Cosmic Gamma-Ray Bursts},
  journal = {The Astrophysical Journal Supplement Series},
  year = {2013},
  volume = {207},
  number = {2},
  pages = {39},
  doi = {10.1088/0067-0049/207/2/39}
}

@article{Palshin2013,
  author = {Pal'shin, V. and others},
  title = {Interplanetary Network Localizations of Konus Short Gamma-Ray Bursts},
  journal = {The Astrophysical Journal Supplement Series},
  year = {2013},
  volume = {207},
  number = {2},
  pages = {38},
  doi = {10.1088/0067-0049/207/2/38}
}

@misc{HEASARCIPN,
  author = {{HEASARC}},
  title = {IPNGRB -- Gamma-Ray Bursts from the Interplanetary Network},
  year = {2025},
  howpublished = {\url{https://heasarc.gsfc.nasa.gov/w3browse/all/ipngrb.html}}
}

@article{Cash:1979vz,
  author         = "Cash, W.",
  title          = "{Parameter estimation in astronomy through application of the likelihood ratio}",
  journal        = "Astrophys. J.",
  volume         = "228",
  pages          = "939--947",
  year           = "1979",
  doi            = "10.1086/156922"
}

@article{Carr:2016drx,
  author       = {B. Carr, F. Kühnel and M. Sandstad},
  title        = {Primordial Black Holes as Dark Matter},
  journal      = {Phys. Rev. D},
  volume       = {94},
  pages        = {083504},
  year         = {2016},
  doi          = {10.1103/PhysRevD.94.083504},
  eprint       = {1607.06077},
  archivePrefix= {arXiv},
  primaryClass = {astro-ph.CO}
}

@article{Carr:2020gox,
  author       = {B. Carr and F. Kühnel},
  title        = {Primordial Black Holes as Dark Matter Candidates},
  journal      = {Annual Review of Nuclear and Particle Science},
  volume       = {70},
  pages        = {355--394},
  year         = {2020},
  doi          = {10.1146/annurev-nucl-050520-125911},
  eprint       = {2006.02838},
  archivePrefix= {arXiv},
  primaryClass = {astro-ph.CO}
}

@article{Villanueva-Domingo:2021spv,
  author       = {P. Villanueva-Domingo and O. Mena and S. Palomares-Ruiz},
  title        = {A brief review on primordial black holes as dark matter},
  journal      = {Frontiers in Astronomy and Space Sciences},
  volume       = {8},
  pages        = {87},
  year         = {2021},
  doi          = {10.3389/fspas.2021.681084},
  eprint       = {2103.12087},
  archivePrefix= {arXiv},
  primaryClass = {astro-ph.CO}
}

@article{Abdo:2015osa,
    author = "Abdo, A. A. and others",
    title = "{Milagro Limits and HAWC Sensitivity for the Rate-Density of Evaporating Primordial Black Holes}",
    eprint = "1407.1686",
    archivePrefix = "arXiv",
    primaryClass = "astro-ph.HE",
    doi = "10.1016/j.astropartphys.2014.10.007",
    journal = "Astropart. Phys.",
    volume = "64",
    pages = "4--12",
    year = "2015"
}

@article{Albert:2019lkt,
    author = "Albert, A. and others",
    collaboration = "HAWC",
    title = "{Constraining the Local Burst Rate Density of Primordial Black Holes with HAWC}",
    eprint = "1911.04356",
    archivePrefix = "arXiv",
    primaryClass = "astro-ph.HE",
    doi = "10.1088/1475-7516/2020/04/026",
    journal = "JCAP",
    volume = "04",
    pages = "026",
    year = "2020"
}

@article{HESS:2021pbh,
    author = "Aharonian, F. and others",
    collaboration = "H.E.S.S.",
    title = "{Search for the evaporation of primordial black holes with H.E.S.S.}",
    eprint = "2303.12855",
    archivePrefix = "arXiv",
    primaryClass = "astro-ph.HE",
    doi = "10.1088/1475-7516/2023/04/040",
    journal = "JCAP",
    volume = "04",
    pages = "040",
    year = "2023"
}

@article{FermiLAT:2018pbh,
    author = "Ackermann, M. and others",
    collaboration = "Fermi-LAT",
    title = "{Search for Gamma-Ray Emission from Local Primordial Black Holes with the Fermi Large Area Telescope}",
    eprint = "1802.00100",
    archivePrefix = "arXiv",
    primaryClass = "astro-ph.HE",
    doi = "10.3847/1538-4357/aaac7b",
    journal = "Astrophys. J.",
    volume = "857",
    number = "1",
    pages = "49",
    year = "2018"
}

@article{Baker:2021sno,
    author = "Baker, Michael J. and Thamm, Andrea",
    title = "{Probing the particle spectrum of nature with evaporating black holes}",
    eprint = "2105.10506",
    archivePrefix = "arXiv",
    primaryClass = "hep-ph",
    doi = "10.21468/SciPostPhys.12.5.150",
    journal = "SciPost Phys.",
    volume = "12",
    number = "5",
    pages = "150",
    year = "2022"
}

@article{Arbey:2019cmf,
    author = "Arbey, Alexandre and Auffinger, J{\'e}r{\'e}my",
    title = "{Physics Beyond the Standard Model with BlackHawk v2.0}",
    eprint = "2108.02737",
    archivePrefix = "arXiv",
    primaryClass = "gr-qc",
    reportNumber = "CERN-TH-2021-117",
    doi = "10.1140/epjc/s10052-021-09702-8",
    journal = "Eur. Phys. J. C",
    volume = "81",
    pages = "910",
    year = "2021"
}

@article{Cline:1995,
    author = "Cline, D. B. and Sanders, D. A. and Hong, W.",
    title = "{Further evidence for gamma-ray bursts consistent with primordial black hole evaporation}",
    reportNumber = "UCLA-APH-0081-12-95",
    doi = "10.1086/304480",
    journal = "Astrophys. J.",
    volume = "486",
    pages = "169--178",
    year = "1997"
}

@article{Cline:2007pbh,
    author = "Cline, David B. and Otwinowski, Stan",
    title = "{Evidence for Primordial Black Hole Final Evaporation: Swift, BATSE and KONUS and Comparisons of VSGRBs and Observations of VSB That Have PBH Time Signatures}",
    eprint = "0908.1352",
    archivePrefix = "arXiv",
    primaryClass = "astro-ph.CO",
    month = "8",
    year = "2009",
  journal      = ""      
}

@article{Belyanin:1996,
    author = "Belyanin, A. A. and Kocharovsky, V. V. and Kocharovsky, VI. V.",
    title = "{Gamma-ray bursts from evaporating primordial black holes}",
    doi = "10.1007/BF02676709",
    journal = "Radiophys. Quant. Electron.",
    volume = "41",
    number = "1",
    pages = "22--27",
    year = "1998"
}

@article{Belyanin:1998,
title = {Gamma-ray bursts from the final stage of primordial black hole evaporations},
journal = {Advances in Space Research},
volume = {22},
number = {7},
pages = {1111-1114},
year = {1998},
issn = {0273-1177},
doi = {https://doi.org/10.1016/S0273-1177(98)00204-X},
url = {https://www.sciencedirect.com/science/article/pii/S027311779800204X},
author = {A.A. Belyanin and V.V. Kocharovsky and Vl.V. Kocharovsky}
}

@article{Cline:1996,
  title={Possibility of Unique Detection of Primordial Black Hole Gamma-Ray Bursts},
  author={David B. Cline and Woopyo Hong},
  journal={Astrophysical Journal},
  year={1992},
  volume={401},
  url={https://api.semanticscholar.org/CorpusID:123137864}
}

@article{Page1976,
  author       = {D. N. Page},
  title        = {Particle emission rates from a black hole. {II}. {M}assless particles from a rotating hole},
  journal      = {Phys. Rev. D},
  volume       = {14},
  pages        = {3260--3273},
  year         = {1976},
  doi          = {10.1103/PhysRevD.14.3260}
}

@article{MacGibbonWebber1990,
  author       = {J. H. MacGibbon and B. R. Webber},
  title        = {Quark- and gluon-jet emission from primordial black holes: {T}he instantaneous spectra},
  journal      = {Phys. Rev. D},
  volume       = {41},
  pages        = {3052--3079},
  year         = {1990},
  doi          = {10.1103/PhysRevD.41.3052}
}

@article{Heckler1997,
  author       = {A. F. Heckler},
  title        = {On the formation of a photosphere around evaporating black holes},
  journal      = {Phys. Rev. D},
  volume       = {55},
  pages        = {480--492},
  year         = {1997},
  doi          = {10.1103/PhysRevD.55.480}
}

@article{Kocevski2003,
    author = "Ramirez-Ruiz, E. and Fenimore, E. E.",
    title = "{Pulse width evolution in grbs: evidence for internal shocks}",
    eprint = "astro-ph/9910273",
    archivePrefix = "arXiv",
    reportNumber = "LA-UR-99-5132",
    doi = "10.1086/309260",
    journal = "Astrophys. J.",
    volume = "539",
    pages = "712",
    year = "2000"
}

@article{RydeSvensson2002,
    author = "Ryde, Felix and Pe'er, Asaf",
    title = "{Quasi-blackbody component and radiative efficiency of the prompt emission of gamma-ray bursts}",
    eprint = "0811.4135",
    archivePrefix = "arXiv",
    primaryClass = "astro-ph",
    doi = "10.1088/0004-637X/702/2/1211",
    journal = "Astrophys. J.",
    volume = "702",
    pages = "1211--1229",
    year = "2009"
}

@article{Federico:2024fyt,
    author = "Federico, Kevin and Profumo, Stefano",
    title = "{Black hole explosions as probes of new physics}",
    eprint = "2411.17038",
    archivePrefix = "arXiv",
    primaryClass = "hep-ph",
    doi = "10.1103/PhysRevD.111.063006",
    journal = "Phys. Rev. D",
    volume = "111",
    number = "6",
    pages = "063006",
    year = "2025"
}

@article{Boluna:2023jlo,
    author = "Boluna, Xavier and Profumo, Stefano and Bl{\'e}, Juliette and Hennings, Dana",
    title = "{Searching for Exploding black holes}",
    eprint = "2307.06467",
    archivePrefix = "arXiv",
    primaryClass = "astro-ph.HE",
    doi = "10.1088/1475-7516/2024/04/024",
    journal = "JCAP",
    volume = "04",
    pages = "024",
    year = "2024"
}

@article{Sjostrand:2014zea,
    author = {Sj{\"o}strand, Torbj{\"o}rn and Ask, Stefan and Christiansen, Jesper R. and Corke, Richard and Desai, Nishita and Ilten, Philip and Mrenna, Stephen and Prestel, Stefan and Rasmussen, Christine O. and Skands, Peter Z.},
    title = "{An introduction to PYTHIA 8.2}",
    eprint = "1410.3012",
    archivePrefix = "arXiv",
    primaryClass = "hep-ph",
    reportNumber = "LU-TP-14-36, MCNET-14-22, CERN-PH-TH-2014-190, FERMILAB-PUB-14-316-CD, DESY-14-178, SLAC-PUB-16122",
    doi = "10.1016/j.cpc.2015.01.024",
    journal = "Comput. Phys. Commun.",
    volume = "191",
    pages = "159--177",
    year = "2015"
}

@article{Carr:2016hva,
    author = "Carr, B. J. and Kohri, Kazunori and Sendouda, Yuuiti and Yokoyama, Jun'ichi",
    title = "{Constraints on primordial black holes from the Galactic gamma-ray background}",
    eprint = "1604.05349",
    archivePrefix = "arXiv",
    primaryClass = "astro-ph.CO",
    reportNumber = "RESCEU-16-16, KEK-TH-1895, KEK-COSMO-193",
    doi = "10.1103/PhysRevD.94.044029",
    journal = "Phys. Rev. D",
    volume = "94",
    number = "4",
    pages = "044029",
    year = "2016"
}

@article{Korwar:2023kpy,
    author = "Korwar, Mrunal and Profumo, Stefano",
    title = "{Updated constraints on primordial black hole evaporation}",
    eprint = "2302.04408",
    archivePrefix = "arXiv",
    primaryClass = "hep-ph",
    doi = "10.1088/1475-7516/2023/05/054",
    journal = "JCAP",
    volume = "05",
    pages = "054",
    year = "2023"
}

@article{Korwar:2024ofe,
    author = "Korwar, Mrunal and Profumo, Stefano",
    title = "{Late-forming black holes and the antiproton, gamma-ray, and antihelium excesses}",
    eprint = "2403.18656",
    archivePrefix = "arXiv",
    primaryClass = "astro-ph.CO",
    reportNumber = "N3AS-24-012",
    doi = "10.1103/PhysRevD.111.023032",
    journal = "Phys. Rev. D",
    volume = "111",
    number = "2",
    pages = "023032",
    year = "2025"
}

@article{Ewasiuk:2024ctc,
    author = "Ewasiuk, Chris and Profumo, Stefano",
    title = "{Constraints on the maximal number of dark degrees of freedom from black hole evaporation, cosmic rays, colliders, and supernovae}",
    eprint = "2409.11359",
    archivePrefix = "arXiv",
    primaryClass = "hep-ph",
    doi = "10.1103/PhysRevD.111.015008",
    journal = "Phys. Rev. D",
    volume = "111",
    number = "1",
    pages = "015008",
    year = "2025"
}

\end{document}